%% file: unpred_article_e.tex
\documentclass[12pt]{article}
\usepackage[T2A]{fontenc}
\usepackage[koi8-r]{inputenc} 
\usepackage[american]{babel} 
\usepackage{graphics} 
\begin{document}
\title{Experiments on predictability of word in context \\ and information rate in natural language\footnote{This is a somewhat expanded version of the paper
    Manin, D.Yu. 2006. {\it Experiments on predictability of word in context and information rate in natural language.} J. Information Processes (electronic publication, http://www.jip.ru), 6 (3), 229-236}}
\author{D.~Yu.~Manin}
\maketitle
\abstract{Based on data from a large-scale experiment with human
  subjects, we conclude that the logarithm of probability to guess a word in
  context (unpredictability) depends linearly on the word length. This
  result holds both for poetry and prose, even though with prose, the
  subjects don't know the length of the omitted
  word. We hypothesize that this effect reflects a tendency of natural
  language to have an even information rate.}
\par

\section[*]{Introduction} 

In this paper we report a particular result of an experimental study
on predictability of words in context. The experiment's primary
motivation is the study of some aspects of poetry perception, but the
result reported here is, in the author's view, of a general linguistic
interest. 

The first study of natural text predictability was performed by the
founder of information theory, C.~E.~Shannon \cite{Shan51}. (We'll
note that even in his groundbreaking work \cite{Shan48},
Shannon briefly touched on the relationship between literary qualities
and redundancy by contrasting highly redundant Basic English with
Joyce's ``Finnegan's Wake'' which ``enlarges the vocabulary and is alleged
to achieve a compression of semantic content''.) Shannon presented his
subject with random passages from Jefferson's biography and had her
guess the next letter until the correct guess was recorded. The number
of guesses for each letter was then used to calculate upper and lower
bounds for the entropy of English, which turned out to be between 0.6
and 1.3 bits per character (bpc), much lower than that of a random mix of
the same letters. Shannon's results also indicated that conditional
entropy decreases as more and more text history becomes known to the
subject, up to at least 100 letters. 

Several authors repeated Shannon's experiments with some
modifications. Burton and Licklider \cite{BurLick55} used 10 different texts of similar
style, and fragment lengths of 1, 2, 4, ..., 128, 1000
characters. Their conclusion was that, contrary to Shannon, increasing
history doesn't affect measured entropy when history length exceeds 32
characters.

F\'onagy \cite{Fon} compared predictability of the next
letter for three types of text: poetry, newspaper, and ``a
conversation of two young girls''. Apparently, his technique involved
only one guess per letter, so entropy estimates could not be
calculated (see below), and results are presented in terms of the
rate of correct answers, poetry being much less predictable than both
other types. 

Kolmogorov reported the results of 0.9--1.4 bpc for
Russian texts in his work \cite{Kolm65} that laid the ground of
algorithmic complexity theory. The paper contains no
details on the very ingenious experimental techniue, but it is
described in the well-known monograph by Yaglom \& Yaglom \cite{Yaglom2}. 

Cover and King \cite{CK78} modified Shannon's technique by having
their subjects place bets on the next letter. They showed that the
optimal betting policy would be to distribute available capital among
the possible outcomes according to their probability and so if the
subjects play in an optimal way (which is not self-evident though),
the letter probabilities could be inferred from their bets. Their
estimate of the entropy of English was calculated at 1.3 bpc. This
work also contains an extensive bibliography.

Moradi {\it et al} \cite{Moradi98} first used two different texts (a
textbook on digital signal processing and a novel by Judith Krantz) to
confirm Burton and Licklider's results on the critical history length
(32 characters), then added two more texts (``101 Dalmatians'' and a
federal aviation manual) to study the dependence of entropy on text
type and subject (with somewhat inconclusive results). 

A number of works were devoted to estimating entropy of natural
language by means of statistical analysis, without using human
subjects. One of the first attempts is reported in \cite{Paisley66},
where 39 English translations of 9 classical Greek texts were used to
study entropy dependency on subject matter, style, and period. A very
crude entropy estimate by letter digram frequency was used. For some
of the more recent developments, see \cite{BrownEtAl92},
\cite{Teahan96} and references therein. By the very nature of these
methods they can't utilize meaning (and even syntax) of the text, but
by the brute force of contemporary computers they begin achieving
results that come reasonably close to those demonstrated by human
language speakers.

Our experimental setup differs from the previous work in two important
aspects. First, we have subjects guess whole words, and not individual
characters. Second, the words to be guessed come (generally speaking)
from the middle of a context, rather than at the end of a fragment. In
addition to filling blanks, we present the subjects with two other
task types where authenticity of a presented word is to be
assessed. The reason for this is that while most of the previous
studies were eventually aimed at efficient text compression, we are
interested in literary (chiefly, poetic) texts as works of literature,
and not as mere character strings subject to application of
compression algorithms\footnote{It should be noted though that
efficient compression is important not only {\it per se}, but also for
cryptographic applications as pointed out in \cite{Teahan96}. In
addition, language models developed for the purpose of compression are
successfully used in applications like speech recognition and OCR,
allowing to disambiguate difficult cases and correct errors.}. Our
goal in designing the experiment was to provide researchers in the
field of poetics with hard data to ground some hypotheses that
otherwise are unavoidably speculative. Guessing the next word in
sequence is not the best way to treat literary text, because even an
ordinary sentence like this one is not essentially a linear sequence
of words or characters, but a complex structure with word associations
running all over the place, both forward and backward. A poem, even
more so, is a structure with strongly coordinated parts, which is not
read sequentially, much less written sequentially. Also, practice
shows that even when guessing letter by letter, people almost always
base their next character choice on a tentative word guess. This is
why guessing whole words in context was more appropriate for our
purpose.

However, the results we present here, as already mentioned, are not
relevant to poetics proper, so we will not dwell on this further, and
refer the interested reader to \cite{LM1}. 

\section[*]{Experimental setup}

In their Introduction to the special issue on computational
linguistics using large corpora, Church and Mercer
\cite{ChurchMercer93} note that ``The 1990s have witnessed a
resurgence of interest in 1950s-style empirical and statistical methods
of language analysis''. They attribute this empirical renaissance
primarily to the availability of processing power and of massive
quantities of data. Of course, these factors favor statistical
analysis of texts as character strings. However, wide availability of
computer networks and interactive Web technologies also made it
possible to set up large-scale experiments with human subjects. 

The experiment has the form of an online literary game in
Russian\footnote{http://ygrec.msk.ru}. However,
the players are also fully aware of the research side, have free
access to theoretical background and current experimental results, and
can participate in online discussions. The players are presented with
text fragments in which one of the words is replaced with blanks or with a
different word. Any sequence of 5 or more Cyrillic letters surrounded
by non-letters was considered a ``word''. Words are selected from
fragments randomly. There are three different trial types:

\begin{description}
\item{type~1:} a word is omitted, and is to be guessed.
\item{type~2:} a word is highlighted, and the task is to determine
  whether it is original or replaced.
\item{type~3:} two words are displayed, and the subject has to
  determine which one is the original word.
\end{description}

Incorrect guesses from trials of type~1 are used as replacements in
trials of types 2 and 3. 

Texts are randomly drawn from a corpus of 3439 fragments of mostly
poetic works in a wide range of styles and periods: from Avantgarde to
mass culture and from 18th century to contemporary. Three prosaic
texts are also included (two classic novels, and a contemporary
political essay).

As of this writing, the experiment has been running almost
continuously for three years. Over 8000 people took part
in it and collectively made almost 900,000 guesses, about a third of
which is of type~1. The traditional laboratory experiment could have
never achieved this scale. Of course, the technique has its own
drawbacks, which are discussed in detail in \cite{LM1}. But they are a
small price to pay for statistical relevance, especially if it can't be
achieved in any other way.

\section[*]{Results} 

The specific goal of the experiment is to discover and analyze
systematic differences between different categories of texts from the
viewpoint of how easy it is to a) reconstruct an omitted word, and b)
distinguish the original word from a replacement. However here we'll
consider a particular property of the texts that turns out to be
independent of the text type and so probably characterizes the
language itself rather than specific texts. This property is the
dependency of word unpredictability on its length. 

We define {\it unpredictability} $U$ as the negative binary logarithm
of the probability to guess a word, $U = -\log_2 p_1$, where $p_1$ is
the average rate of correct answers to trials of type~1. For a single
word, this is formally equivalent to Shannon's definition of entropy, $H$. However,
when multiple words are taken into account, entropy should be
calculated as the average logarithm of probability, and not as
the logarithm of average probability, 

\begin{equation}
H = -\frac {1}{N}\sum_{i=1}^N \log_2{p_1^i}
\end{equation}
\begin{equation}
U = -\log_2{\frac{1}{N}\sum_{i=1}^N{p_1^i}}
\end{equation}

Indeed, the logarithm of probability to guess a word equals the
amount of information in bits required to determine the word
choice. Thus, it is this quantity that is subject to averaging. When
dealing with experimental data, it is customary to use frequencies as
estimates of unobservable probabilities. However, there are always
words that were never guessed correctly and have $p_1 = 0$ for which
logarithm is undefined (this is why Shannon's techinque involves
repeated guessing of the same letter until the correct answer is
obtained). Formally, if there is one element in the sequence with zero
(very small, in fact) probability of being guessed, then the amount of
information of the whole sequence may be determined solely by this one
element.

On the other hand, unpredictability as defined above is not sensitive
to the exact probability to guess such words, but only on how many
there are of them. While entropy characterizes the number of tries
required to guess a randomly selected word, unpredictability
characterizes the portion of words that would be guessed on the first
try. They are equal, of course, if all words have the same entropy. 

One way around the problem presented by never-guessed words would be
to assign some arbitrary finite entropy to them. We compared
unpredictability with entropy calculated under this approximation with
two values of the constant: 10 bits (corresponding roughly to wild
guessing using a frequency dictionary) and 3 bits (the low bound). In
both cases, while $H$ is not equal numerically to $U$, they turned out
to be in an almost monotonic, approximately linear
correspondence. This probably means that the fraction of hard-to-guess
words co-varies with unpredictability of the rest of the
words. Because of this, we prefer to work in terms of
unpredictability, rather than introducing arbitrary hypotheses to
calculate an entropy value of dubious validity. 

Unpredictability as a function of word length calculated over all
words of the same length across all texts is plotted in Fig. 1 and
Fig. 2 (where word length is measured in characters and syllables
respectively). Confidence intervals on the graphs are calculated based
on the standard deviation of the binomial distribution (since the data
comes from a series of independent trials with two possible outcomes
in each: a guess may be correct or incorrect). 

\begin{figure}[htp]
\centering
\input{nepred_all_let_e.tex}
\caption{Unpredictability as a function of word length in characters,
  all texts}\label{fig:1}
\end{figure}
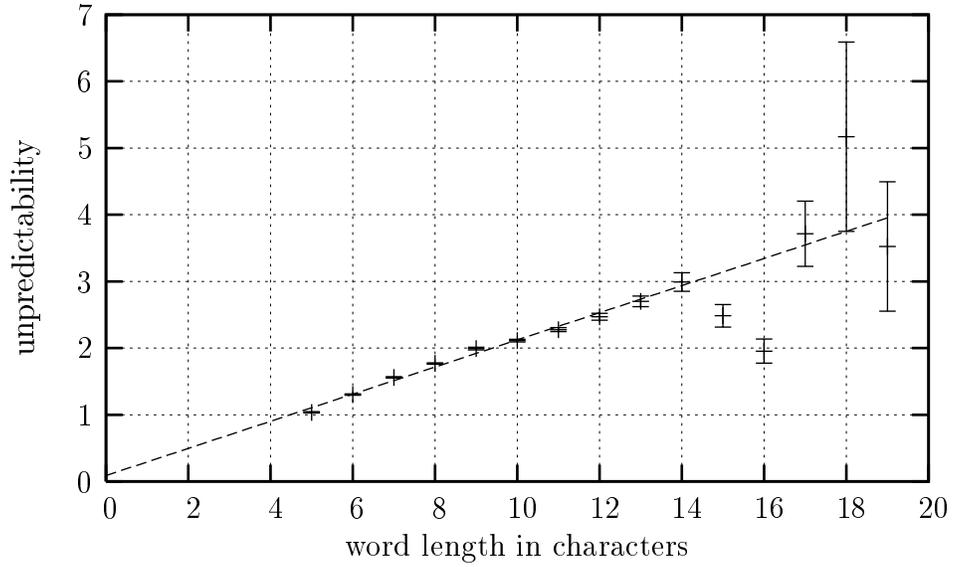

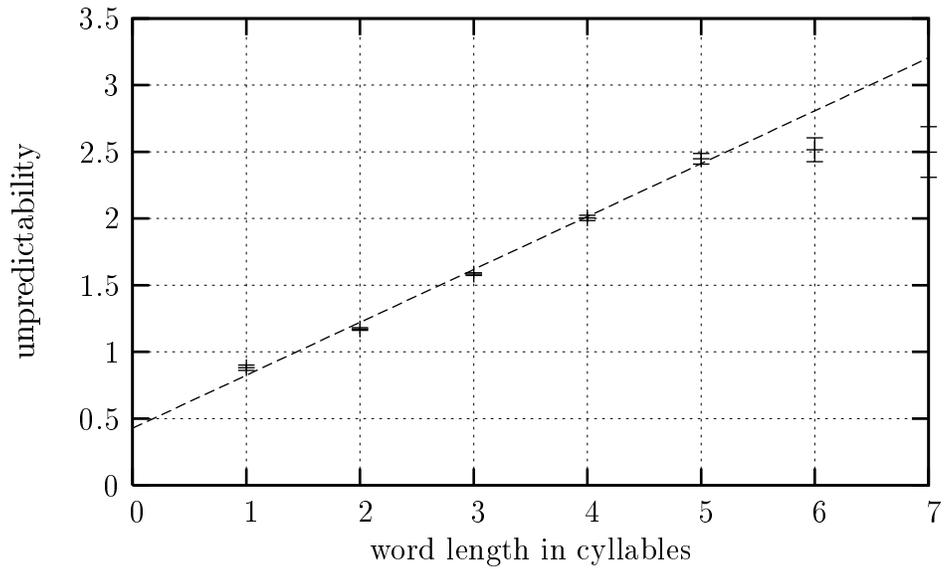
\begin{figure}[htp]
\centering
\input{nepred_all_cyl_e.tex}
\caption{Unpredictability as a function of word length in syllables,
  all texts}\label{fig:2}
\end{figure}

In the range from 5 to 14 characters and from 1 to 5 syllables, an
excellent linear dependence is observed. Longer words are rare, so
the data for them is significantly less statistically reliable. We'll only discuss
the linear dependence in the range where it is definitely valid. 

\section[*]{Discussion} 

It is very difficult, for the reasons mentioned above, to compare our
results with previous studies. However, there are two points of
comparison that can be made. First, we can roughly estimate the effect
of word guessing in context as opposed to guessing the next word in
sequence. Recall that Shannon \cite{Shan51} estimated zeroth-order
word entropy for English based on Zipf's law to be 11.82 bits per word
(bpw). Brown {\it et al} \cite{BrownEtAl92} used a word trigram model to
achieve an entropy estimate of 1.72 bpc, which translates to 7.74 bpw for
average word length of 4.5 characters in English. This means that
trigram word probabilities contribute $11.82-7.74=4.08$ bpw for
prediction of word in sequence. But word in context participates in
three trigrams at once: as the last, the middle and the first word of
a trigram. Only the first trigram is available when the model is
predicting the next word, but all three trigrams could be used to fill
in an omitted word (this is a hypothetical experiment which was not
actually performed). Of course, they are not statistically
independent, and as a rough estimate we can assume that the last
trigram contributes somewhat less information than the first one,
while the middle trigram contributes very little (since all of its
words are already accounted for). In other words, we could expect this
model to have about 4 bpw more information when guessing words in
context, which is very significant. 

The second point of comparison is provided by \cite{SG96} (Fig. 13
there), where entropy is plotted for the $n$-th letter of each word,
versus its position $n$. Entropy was estimated using a Ziv--Lempel
type algorithm. It is well-known that guessing is least confident
at the word boundaries for both human subjects and computer
algorithms, and this chart quantifies the observation: the first
letter has the entropy of 4 bpc, which drops quickly to about 0.6--0.7~bpc for
the 5th letter and then stays surprizingly constant all the way
through the 16th character. This chart is practically the same for the
original text and a text with randomly permuted words, which gives a
telling evidence of the current language models' strengths and
weaknesses. For the purposes of this discussion, the data allows to
reconstruct the dependency of word entropy on the word length as
$h_n^{(w)} = \sum_{i=1}^n{h_i^{(l)}}$, where $h_n^{(w)}$ is the entropy
of words of length $n$, and $h_i^{(l)}$ is the entropy of the $i$-th
letter in a word. This dependency, valid for the language model in
\cite{SG96}, has a steep increase from 1 through 5 characters, and
then an approximately linear growth with a much shallower slope of
0.6--0.7~bpc. This is very different from our Fig.~1, and even though
our data is on unpredictability, rather than entropy, the difference
is probably significant.

In fact, our result may at first glance seem trivial. Indeed,
according to a theorem due to Shannon (Theorem 3 in \cite{Shan48}),
for a character sequence emitted by a stationary ergodic source,
almost all subsequences of length $n$ have the same probability
exponential in $n$: $P_n = 2^{-Hn}$ for large enough length ($H$
is the entropy of the source). However,
this explanation is not valid here for several reasons. Even if we set
aside the question of natural language ergodicity, from the formal
point of view, the theorem requires that $n$ is large enough so that
all possible letter digrams are likely to be encountered more than
once (many times, in fact). Needless to say that the length of a
single word is much less than that. Practically, if this explanation
were to be adopted, we'd expect the probability to guess a word to be
on the order $P_n$, which is much smaller than the observed
probability. In fact, the only reason our subjects are able to guess
words in context is that the words are connected to the context and
make sense in it, while under the assumptions of Shannon's theorem,
the equiprobable subsequences are asymptotically independent of the
context.

Another tentative argument is to presume that the total number of
words in the language (either in the vocabulary or in texts, which is
not the same thing) of a given length increases with length, which
makes longer words harder to guess due to sheer expansion of
possibilities. If there had been exponential expansion of vocabulary
with word length, we could argue that contextual restrictions on word
choice cut the number of choices by a constant factor (on the
average), so the number of words satisfying these restrictions still
grows exponentially with word length. However, the data does not support
this idea. Distribution of words by length, whether computed from the
actual texts or from a dictionary (we used a Russian frequency dictionary
containing 32000 words \cite{FreqDict}), is not even monotonic, let
alone exponentially growing. The number of different words grows up to
about 8 characters of length, then decreases. This behavior is in no
way reflected in Figs 1, 2, so we can conclude that the total number
of dictionary words of a given length is not a factor in guessing success. 

In fact, the word length distribution could have had a direct effect
on unpredictability only if the word length were known to the
subject. But this is generally not the case. Subjects in our
experiment are not given any external clue as to the length of the
omitted word. Since Russian verse is for the most part metric, the
syllabic length of a line is typically known, and this allows to
predict the syllabic length of the omitted word with a great deal of
certainty. However unpredictability depends on word length in exactly
the same way for poetry and prose (see Fig. 3), and in prose there are
no external {\it or} internal clues for the word length. \footnote{It
is also worth noting that average unpredictability of words in poetry
and prose is surprisingly close. In poetry, it turns out,
predictability due to meter and rhyme is counteracted by increased
unpredictability of semantics and, possibly, grammar. Notably, these
two tendencies almost balance each other. This phenomenon and its
significance is discussed at length in  \cite{LM1}.}
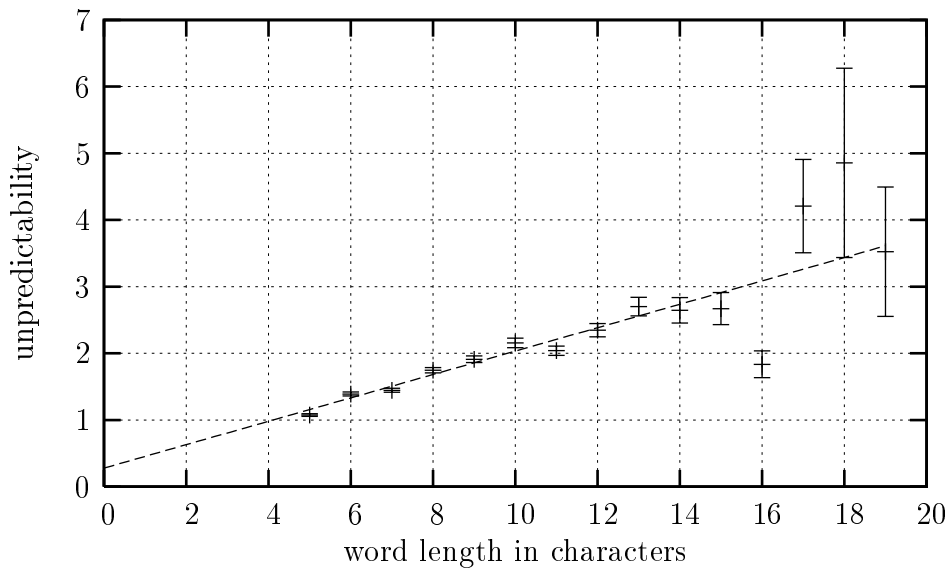
\begin{figure}[htp]
\centering
\input{nepred_prose_let_e.tex}
\caption{Unpredictability as a function of word length in characters,
  prose only}\label{fig:3}
\end{figure}

This leaves us with the only reasonable explanation for the observed
dependency: in course of its evolution, the language tends to even out
information rate, so that longer words carry proportionally more
information. This would be a natural assumption, since an uneven
information rate is inefficient: some portions will underutilize the
bandwidth of the channel, and some will overutilize it and diminish
error-correction capabilities. In other words, as language changes
over time, some words and grammatical forms that are too long will be
shortened, and those that are too short will be expanded and
reinforced.

It is interesting to note that this hypothesis was also proposed in
passing by Church and Mercer in a different context in
\cite{ChurchMercer93}. Discussing applications of trigram
word-prediction models to speech recognition, they write (page 12):

\begin{quotation}
In general, high-frequency function words like {\it to} and {\it the}, which are
acoustically short, are more predictable than content words like
{\it resolve} and {\it important}, which are longer. This is convenient for speech
recognition because it means that the language model provides more
powerful constraints just when the acoustic model is having the
toughest time. One suspects that this is not an accident, but rather a
natural result of the evolution of speech to fill the human needs for
reliable communication in the presence of noise.
\end{quotation}

A feature that is ``convenient for speech recognition'' is, indeed,
not to be unexpected in natural language, and from our results it
appears that its extent is much broader than could be suggested by
Church and Mercer's observation. Of course, this is only one of many
mechanisms that drive language change, and it only acts statistically,
so any given language state will have low-redundancy and
high-redundancy pockets. Thus, any Russian speaker knows how difficult
it is to distinguish between {\it mne nado} 'I need' and {\it ne
nado} 'please don't'. Moreover, it is likely that this change
typically proceeds by vacillations. As an example consider the
evolution of negation in English according to \cite{HockJoseph} (p. 175--176):
\begin{quotation}
the original Old English word of negation was {\it ne}, as in {\it ic ne w\=at}, 'I don't know'. This ordinary mode of negation could be
  reinforced by the hyperbolic use of either {\it wiht} 'something,
  anything' or {\it n\=awiht} 'nothing, not anything' [...]. As
  time progressed, the hyperbolic force of {\it (n\=a)wiht} began
  to fade [...] and the form {\it n\=awiht} came to be interpreted
  as part of a two-part, ``discontinuous'' marker of negation {\it ne
  ... n\=awiht} [...]. But once ordinary negation was expressed by
  two words, {\it ne} and {\it n\=awiht}, the stage was set for
  ellipsis to come in and to eliminate the seeming redundancy. The
  result was that {\it ne}, the word that originally had been the
  marker of negation, was deleted, and {\it not}, the reflex of
  originally hyperbolic {\it n\=awiht} became the only marker of
  negation. [...] (Modern English has introduced further changes
  through the introduction of the ``helping word'' {\it do}.)
\end{quotation}

This looks very much like oscillations resulting from an iterative
search for the optimum length of a particular grammatical form. It's
all the more amazing then, how this tendency, despite its statistical
and non-stationary character, beautifully manifests itself in the
data.

\medskip

{\bf Addendum.} After this paper was published in {\it J. Information
  Proc.}, the author became aware of the following works in
  which effects of the same nature was discovered on different levels:

{\it The discource level.} Genzel and Charniak \cite{GenzelCharniak}
study the entropy of a sentence depending on its position in the
text. They show that the entropy calculated by a language model, which
does not account for semantics, increases somewhat in the initial
portion of the text. They conclude that the hypothetical entropy value
with account for semantics would be constant, because the content of
the preceding text would help predicting the following text. 

{\it The sentence level.} The authors of \cite{Jaeger06} consider the
English sentences with optional relativizer {\it that}. They
demonstrate experimentally that the speakers tend to utter the
optional relativizer more frequently in those sentences where
information density is higher thus ``diluting'' them. This can be
interpreted as a tendency to homogenize information density.

{\it The syllabe level} Aulett and Turk \cite{AylettTurk} demonstrate
that in spontaneous speech, those syllables that are less predictable,
are increased in duration and prosodic prominence. In this way,
speakers tend to smooth out the redundancy level. 

\medskip

{\bf Acknowledgements.} I am grateful to D.~Flitman for hosting the
website, to Yu.~Manin, M.~Verbitsky, Yu.~Fridman, R.~Leibov, G.~Mints
and many others for fruitful discussions. This work could not have
happened without the generous support of over 8000 people who
generated experimental data by playing the game.


\end{document}

%% file: nepred_all_let_e.tex
\begingroup%
  \makeatletter%
  \newcommand{\GNUPLOTspecial}{%
    \@sanitize\catcode`\%=14\relax\special}%
  \setlength{\unitlength}{0.1bp}%
\begin{picture}(3600,2160)(0,0)%
{\GNUPLOTspecial{"
/gnudict 256 dict def
gnudict begin
/Color false def
/Solid false def
/gnulinewidth 5.000 def
/userlinewidth gnulinewidth def
/vshift -33 def
/dl {10.0 mul} def
/hpt_ 31.5 def
/vpt_ 31.5 def
/hpt hpt_ def
/vpt vpt_ def
/Rounded false def
/M {moveto} bind def
/L {lineto} bind def
/R {rmoveto} bind def
/V {rlineto} bind def
/N {newpath moveto} bind def
/C {setrgbcolor} bind def
/f {rlineto fill} bind def
/vpt2 vpt 2 mul def
/hpt2 hpt 2 mul def
/Lshow { currentpoint stroke M
  0 vshift R show } def
/Rshow { currentpoint stroke M
  dup stringwidth pop neg vshift R show } def
/Cshow { currentpoint stroke M
  dup stringwidth pop -2 div vshift R show } def
/UP { dup vpt_ mul /vpt exch def hpt_ mul /hpt exch def
  /hpt2 hpt 2 mul def /vpt2 vpt 2 mul def } def
/DL { Color {setrgbcolor Solid {pop []} if 0 setdash }
 {pop pop pop 0 setgray Solid {pop []} if 0 setdash} ifelse } def
/BL { stroke userlinewidth 2 mul setlinewidth
      Rounded { 1 setlinejoin 1 setlinecap } if } def
/AL { stroke userlinewidth 2 div setlinewidth
      Rounded { 1 setlinejoin 1 setlinecap } if } def
/UL { dup gnulinewidth mul /userlinewidth exch def
      dup 1 lt {pop 1} if 10 mul /udl exch def } def
/PL { stroke userlinewidth setlinewidth
      Rounded { 1 setlinejoin 1 setlinecap } if } def
/LTw { PL [] 1 setgray } def
/LTb { BL [] 0 0 0 DL } def
/LTa { AL [1 udl mul 2 udl mul] 0 setdash 0 0 0 setrgbcolor } def
/LT0 { PL [] 1 0 0 DL } def
/LT1 { PL [4 dl 2 dl] 0 1 0 DL } def
/LT2 { PL [2 dl 3 dl] 0 0 1 DL } def
/LT3 { PL [1 dl 1.5 dl] 1 0 1 DL } def
/LT4 { PL [5 dl 2 dl 1 dl 2 dl] 0 1 1 DL } def
/LT5 { PL [4 dl 3 dl 1 dl 3 dl] 1 1 0 DL } def
/LT6 { PL [2 dl 2 dl 2 dl 4 dl] 0 0 0 DL } def
/LT7 { PL [2 dl 2 dl 2 dl 2 dl 2 dl 4 dl] 1 0.3 0 DL } def
/LT8 { PL [2 dl 2 dl 2 dl 2 dl 2 dl 2 dl 2 dl 4 dl] 0.5 0.5 0.5 DL } def
/Pnt { stroke [] 0 setdash
   gsave 1 setlinecap M 0 0 V stroke grestore } def
/Dia { stroke [] 0 setdash 2 copy vpt add M
  hpt neg vpt neg V hpt vpt neg V
  hpt vpt V hpt neg vpt V closepath stroke
  Pnt } def
/Pls { stroke [] 0 setdash vpt sub M 0 vpt2 V
  currentpoint stroke M
  hpt neg vpt neg R hpt2 0 V stroke
  } def
/Box { stroke [] 0 setdash 2 copy exch hpt sub exch vpt add M
  0 vpt2 neg V hpt2 0 V 0 vpt2 V
  hpt2 neg 0 V closepath stroke
  Pnt } def
/Crs { stroke [] 0 setdash exch hpt sub exch vpt add M
  hpt2 vpt2 neg V currentpoint stroke M
  hpt2 neg 0 R hpt2 vpt2 V stroke } def
/TriU { stroke [] 0 setdash 2 copy vpt 1.12 mul add M
  hpt neg vpt -1.62 mul V
  hpt 2 mul 0 V
  hpt neg vpt 1.62 mul V closepath stroke
  Pnt  } def
/Star { 2 copy Pls Crs } def
/BoxF { stroke [] 0 setdash exch hpt sub exch vpt add M
  0 vpt2 neg V  hpt2 0 V  0 vpt2 V
  hpt2 neg 0 V  closepath fill } def
/TriUF { stroke [] 0 setdash vpt 1.12 mul add M
  hpt neg vpt -1.62 mul V
  hpt 2 mul 0 V
  hpt neg vpt 1.62 mul V closepath fill } def
/TriD { stroke [] 0 setdash 2 copy vpt 1.12 mul sub M
  hpt neg vpt 1.62 mul V
  hpt 2 mul 0 V
  hpt neg vpt -1.62 mul V closepath stroke
  Pnt  } def
/TriDF { stroke [] 0 setdash vpt 1.12 mul sub M
  hpt neg vpt 1.62 mul V
  hpt 2 mul 0 V
  hpt neg vpt -1.62 mul V closepath fill} def
/DiaF { stroke [] 0 setdash vpt add M
  hpt neg vpt neg V hpt vpt neg V
  hpt vpt V hpt neg vpt V closepath fill } def
/Pent { stroke [] 0 setdash 2 copy gsave
  translate 0 hpt M 4 {72 rotate 0 hpt L} repeat
  closepath stroke grestore Pnt } def
/PentF { stroke [] 0 setdash gsave
  translate 0 hpt M 4 {72 rotate 0 hpt L} repeat
  closepath fill grestore } def
/Circle { stroke [] 0 setdash 2 copy
  hpt 0 360 arc stroke Pnt } def
/CircleF { stroke [] 0 setdash hpt 0 360 arc fill } def
/C0 { BL [] 0 setdash 2 copy moveto vpt 90 450  arc } bind def
/C1 { BL [] 0 setdash 2 copy        moveto
       2 copy  vpt 0 90 arc closepath fill
               vpt 0 360 arc closepath } bind def
/C2 { BL [] 0 setdash 2 copy moveto
       2 copy  vpt 90 180 arc closepath fill
               vpt 0 360 arc closepath } bind def
/C3 { BL [] 0 setdash 2 copy moveto
       2 copy  vpt 0 180 arc closepath fill
               vpt 0 360 arc closepath } bind def
/C4 { BL [] 0 setdash 2 copy moveto
       2 copy  vpt 180 270 arc closepath fill
               vpt 0 360 arc closepath } bind def
/C5 { BL [] 0 setdash 2 copy moveto
       2 copy  vpt 0 90 arc
       2 copy moveto
       2 copy  vpt 180 270 arc closepath fill
               vpt 0 360 arc } bind def
/C6 { BL [] 0 setdash 2 copy moveto
      2 copy  vpt 90 270 arc closepath fill
              vpt 0 360 arc closepath } bind def
/C7 { BL [] 0 setdash 2 copy moveto
      2 copy  vpt 0 270 arc closepath fill
              vpt 0 360 arc closepath } bind def
/C8 { BL [] 0 setdash 2 copy moveto
      2 copy vpt 270 360 arc closepath fill
              vpt 0 360 arc closepath } bind def
/C9 { BL [] 0 setdash 2 copy moveto
      2 copy  vpt 270 450 arc closepath fill
              vpt 0 360 arc closepath } bind def
/C10 { BL [] 0 setdash 2 copy 2 copy moveto vpt 270 360 arc closepath fill
       2 copy moveto
       2 copy vpt 90 180 arc closepath fill
               vpt 0 360 arc closepath } bind def
/C11 { BL [] 0 setdash 2 copy moveto
       2 copy  vpt 0 180 arc closepath fill
       2 copy moveto
       2 copy  vpt 270 360 arc closepath fill
               vpt 0 360 arc closepath } bind def
/C12 { BL [] 0 setdash 2 copy moveto
       2 copy  vpt 180 360 arc closepath fill
               vpt 0 360 arc closepath } bind def
/C13 { BL [] 0 setdash  2 copy moveto
       2 copy  vpt 0 90 arc closepath fill
       2 copy moveto
       2 copy  vpt 180 360 arc closepath fill
               vpt 0 360 arc closepath } bind def
/C14 { BL [] 0 setdash 2 copy moveto
       2 copy  vpt 90 360 arc closepath fill
               vpt 0 360 arc } bind def
/C15 { BL [] 0 setdash 2 copy vpt 0 360 arc closepath fill
               vpt 0 360 arc closepath } bind def
/Rec   { newpath 4 2 roll moveto 1 index 0 rlineto 0 exch rlineto
       neg 0 rlineto closepath } bind def
/Square { dup Rec } bind def
/Bsquare { vpt sub exch vpt sub exch vpt2 Square } bind def
/S0 { BL [] 0 setdash 2 copy moveto 0 vpt rlineto BL Bsquare } bind def
/S1 { BL [] 0 setdash 2 copy vpt Square fill Bsquare } bind def
/S2 { BL [] 0 setdash 2 copy exch vpt sub exch vpt Square fill Bsquare } bind def
/S3 { BL [] 0 setdash 2 copy exch vpt sub exch vpt2 vpt Rec fill Bsquare } bind def
/S4 { BL [] 0 setdash 2 copy exch vpt sub exch vpt sub vpt Square fill Bsquare } bind def
/S5 { BL [] 0 setdash 2 copy 2 copy vpt Square fill
       exch vpt sub exch vpt sub vpt Square fill Bsquare } bind def
/S6 { BL [] 0 setdash 2 copy exch vpt sub exch vpt sub vpt vpt2 Rec fill Bsquare } bind def
/S7 { BL [] 0 setdash 2 copy exch vpt sub exch vpt sub vpt vpt2 Rec fill
       2 copy vpt Square fill
       Bsquare } bind def
/S8 { BL [] 0 setdash 2 copy vpt sub vpt Square fill Bsquare } bind def
/S9 { BL [] 0 setdash 2 copy vpt sub vpt vpt2 Rec fill Bsquare } bind def
/S10 { BL [] 0 setdash 2 copy vpt sub vpt Square fill 2 copy exch vpt sub exch vpt Square fill
       Bsquare } bind def
/S11 { BL [] 0 setdash 2 copy vpt sub vpt Square fill 2 copy exch vpt sub exch vpt2 vpt Rec fill
       Bsquare } bind def
/S12 { BL [] 0 setdash 2 copy exch vpt sub exch vpt sub vpt2 vpt Rec fill Bsquare } bind def
/S13 { BL [] 0 setdash 2 copy exch vpt sub exch vpt sub vpt2 vpt Rec fill
       2 copy vpt Square fill Bsquare } bind def
/S14 { BL [] 0 setdash 2 copy exch vpt sub exch vpt sub vpt2 vpt Rec fill
       2 copy exch vpt sub exch vpt Square fill Bsquare } bind def
/S15 { BL [] 0 setdash 2 copy Bsquare fill Bsquare } bind def
/D0 { gsave translate 45 rotate 0 0 S0 stroke grestore } bind def
/D1 { gsave translate 45 rotate 0 0 S1 stroke grestore } bind def
/D2 { gsave translate 45 rotate 0 0 S2 stroke grestore } bind def
/D3 { gsave translate 45 rotate 0 0 S3 stroke grestore } bind def
/D4 { gsave translate 45 rotate 0 0 S4 stroke grestore } bind def
/D5 { gsave translate 45 rotate 0 0 S5 stroke grestore } bind def
/D6 { gsave translate 45 rotate 0 0 S6 stroke grestore } bind def
/D7 { gsave translate 45 rotate 0 0 S7 stroke grestore } bind def
/D8 { gsave translate 45 rotate 0 0 S8 stroke grestore } bind def
/D9 { gsave translate 45 rotate 0 0 S9 stroke grestore } bind def
/D10 { gsave translate 45 rotate 0 0 S10 stroke grestore } bind def
/D11 { gsave translate 45 rotate 0 0 S11 stroke grestore } bind def
/D12 { gsave translate 45 rotate 0 0 S12 stroke grestore } bind def
/D13 { gsave translate 45 rotate 0 0 S13 stroke grestore } bind def
/D14 { gsave translate 45 rotate 0 0 S14 stroke grestore } bind def
/D15 { gsave translate 45 rotate 0 0 S15 stroke grestore } bind def
/DiaE { stroke [] 0 setdash vpt add M
  hpt neg vpt neg V hpt vpt neg V
  hpt vpt V hpt neg vpt V closepath stroke } def
/BoxE { stroke [] 0 setdash exch hpt sub exch vpt add M
  0 vpt2 neg V hpt2 0 V 0 vpt2 V
  hpt2 neg 0 V closepath stroke } def
/TriUE { stroke [] 0 setdash vpt 1.12 mul add M
  hpt neg vpt -1.62 mul V
  hpt 2 mul 0 V
  hpt neg vpt 1.62 mul V closepath stroke } def
/TriDE { stroke [] 0 setdash vpt 1.12 mul sub M
  hpt neg vpt 1.62 mul V
  hpt 2 mul 0 V
  hpt neg vpt -1.62 mul V closepath stroke } def
/PentE { stroke [] 0 setdash gsave
  translate 0 hpt M 4 {72 rotate 0 hpt L} repeat
  closepath stroke grestore } def
/CircE { stroke [] 0 setdash 
  hpt 0 360 arc stroke } def
/Opaque { gsave closepath 1 setgray fill grestore 0 setgray closepath } def
/DiaW { stroke [] 0 setdash vpt add M
  hpt neg vpt neg V hpt vpt neg V
  hpt vpt V hpt neg vpt V Opaque stroke } def
/BoxW { stroke [] 0 setdash exch hpt sub exch vpt add M
  0 vpt2 neg V hpt2 0 V 0 vpt2 V
  hpt2 neg 0 V Opaque stroke } def
/TriUW { stroke [] 0 setdash vpt 1.12 mul add M
  hpt neg vpt -1.62 mul V
  hpt 2 mul 0 V
  hpt neg vpt 1.62 mul V Opaque stroke } def
/TriDW { stroke [] 0 setdash vpt 1.12 mul sub M
  hpt neg vpt 1.62 mul V
  hpt 2 mul 0 V
  hpt neg vpt -1.62 mul V Opaque stroke } def
/PentW { stroke [] 0 setdash gsave
  translate 0 hpt M 4 {72 rotate 0 hpt L} repeat
  Opaque stroke grestore } def
/CircW { stroke [] 0 setdash 
  hpt 0 360 arc Opaque stroke } def
/BoxFill { gsave Rec 1 setgray fill grestore } def
/BoxColFill {
  gsave Rec
  /Fillden exch def
  currentrgbcolor
  /ColB exch def /ColG exch def /ColR exch def
  /ColR ColR Fillden mul Fillden sub 1 add def
  /ColG ColG Fillden mul Fillden sub 1 add def
  /ColB ColB Fillden mul Fillden sub 1 add def
  ColR ColG ColB setrgbcolor
  fill grestore } def
%
%
/PatternFill { gsave /PFa [ 9 2 roll ] def
    PFa 0 get PFa 2 get 2 div add PFa 1 get PFa 3 get 2 div add translate
    PFa 2 get -2 div PFa 3 get -2 div PFa 2 get PFa 3 get Rec
    gsave 1 setgray fill grestore clip
    currentlinewidth 0.5 mul setlinewidth
    /PFs PFa 2 get dup mul PFa 3 get dup mul add sqrt def
    0 0 M PFa 5 get rotate PFs -2 div dup translate
	0 1 PFs PFa 4 get div 1 add floor cvi
	{ PFa 4 get mul 0 M 0 PFs V } for
    0 PFa 6 get ne {
	0 1 PFs PFa 4 get div 1 add floor cvi
	{ PFa 4 get mul 0 2 1 roll M PFs 0 V } for
    } if
    stroke grestore } def
/Symbol-Oblique /Symbol findfont [1 0 .167 1 0 0] makefont
dup length dict begin {1 index /FID eq {pop pop} {def} ifelse} forall
currentdict end definefont pop
end
gnudict begin
gsave
0 0 translate
0.100 0.100 scale
0 setgray
newpath
1.000 UL
LTb
1.000 UL
LTa
350 300 M
3100 0 V
1.000 UL
LTb
350 300 M
63 0 V
3037 0 R
-63 0 V
1.000 UL
LTb
1.000 UL
LTa
350 551 M
3100 0 V
1.000 UL
LTb
350 551 M
63 0 V
3037 0 R
-63 0 V
1.000 UL
LTb
1.000 UL
LTa
350 803 M
3100 0 V
1.000 UL
LTb
350 803 M
63 0 V
3037 0 R
-63 0 V
1.000 UL
LTb
1.000 UL
LTa
350 1054 M
3100 0 V
1.000 UL
LTb
350 1054 M
63 0 V
3037 0 R
-63 0 V
1.000 UL
LTb
1.000 UL
LTa
350 1306 M
3100 0 V
1.000 UL
LTb
350 1306 M
63 0 V
3037 0 R
-63 0 V
1.000 UL
LTb
1.000 UL
LTa
350 1557 M
3100 0 V
1.000 UL
LTb
350 1557 M
63 0 V
3037 0 R
-63 0 V
1.000 UL
LTb
1.000 UL
LTa
350 1809 M
3100 0 V
1.000 UL
LTb
350 1809 M
63 0 V
3037 0 R
-63 0 V
1.000 UL
LTb
1.000 UL
LTa
350 2060 M
3100 0 V
1.000 UL
LTb
350 2060 M
63 0 V
3037 0 R
-63 0 V
1.000 UL
LTb
1.000 UL
LTa
350 300 M
0 1760 V
1.000 UL
LTb
350 300 M
0 63 V
0 1697 R
0 -63 V
1.000 UL
LTb
1.000 UL
LTa
660 300 M
0 1760 V
1.000 UL
LTb
660 300 M
0 63 V
0 1697 R
0 -63 V
1.000 UL
LTb
1.000 UL
LTa
970 300 M
0 1760 V
1.000 UL
LTb
970 300 M
0 63 V
0 1697 R
0 -63 V
1.000 UL
LTb
1.000 UL
LTa
1280 300 M
0 1760 V
1.000 UL
LTb
1280 300 M
0 63 V
0 1697 R
0 -63 V
1.000 UL
LTb
1.000 UL
LTa
1590 300 M
0 1760 V
1.000 UL
LTb
1590 300 M
0 63 V
0 1697 R
0 -63 V
1.000 UL
LTb
1.000 UL
LTa
1900 300 M
0 1760 V
1.000 UL
LTb
1900 300 M
0 63 V
0 1697 R
0 -63 V
1.000 UL
LTb
1.000 UL
LTa
2210 300 M
0 1760 V
1.000 UL
LTb
2210 300 M
0 63 V
0 1697 R
0 -63 V
1.000 UL
LTb
1.000 UL
LTa
2520 300 M
0 1760 V
1.000 UL
LTb
2520 300 M
0 63 V
0 1697 R
0 -63 V
1.000 UL
LTb
1.000 UL
LTa
2830 300 M
0 1760 V
1.000 UL
LTb
2830 300 M
0 63 V
0 1697 R
0 -63 V
1.000 UL
LTb
1.000 UL
LTa
3140 300 M
0 1760 V
1.000 UL
LTb
3140 300 M
0 63 V
0 1697 R
0 -63 V
1.000 UL
LTb
1.000 UL
LTa
3450 300 M
0 1760 V
1.000 UL
LTb
3450 300 M
0 63 V
0 1697 R
0 -63 V
1.000 UL
LTb
1.000 UL
LTb
350 300 M
3100 0 V
0 1760 V
-3100 0 V
350 300 L
LTb
LTb
1.000 UP
1.000 UP
1.000 UL
LT0
1125 558 M
0 5 V
-31 -5 R
62 0 V
-62 5 R
62 0 V
124 62 R
0 5 V
-31 -5 R
62 0 V
-62 5 R
62 0 V
124 60 R
0 5 V
-31 -5 R
62 0 V
-62 5 R
62 0 V
124 47 R
0 5 V
-31 -5 R
62 0 V
-62 5 R
62 0 V
124 49 R
0 10 V
-31 -10 R
62 0 V
-62 10 R
62 0 V
124 20 R
0 10 V
-31 -10 R
62 0 V
-62 10 R
62 0 V
124 29 R
0 15 V
-31 -15 R
62 0 V
-62 15 R
62 0 V
124 28 R
0 26 V
-31 -26 R
62 0 V
-62 26 R
62 0 V
124 25 R
0 40 V
-31 -40 R
62 0 V
-62 40 R
62 0 V
124 18 R
0 70 V
-31 -70 R
62 0 V
-62 70 R
62 0 V
2675 882 M
0 85 V
-31 -85 R
62 0 V
-62 85 R
62 0 V
2830 746 M
0 91 V
-31 -91 R
62 0 V
-62 91 R
62 0 V
124 274 R
0 246 V
-31 -246 R
62 0 V
-62 246 R
62 0 V
124 -114 R
0 714 V
-31 -714 R
62 0 V
-62 714 R
62 0 V
3295 942 M
0 488 V
3264 942 M
62 0 V
-62 488 R
62 0 V
1125 560 Pls
1280 627 Pls
1435 692 Pls
1590 745 Pls
1745 801 Pls
1900 831 Pls
2055 872 Pls
2210 921 Pls
2365 979 Pls
2520 1052 Pls
2675 925 Pls
2830 791 Pls
2985 1234 Pls
3140 1600 Pls
3295 1186 Pls
1.000 UL
LT1
350 322 M
30 10 V
29 10 V
30 10 V
30 9 V
30 10 V
29 10 V
30 10 V
30 10 V
30 9 V
29 10 V
30 10 V
30 10 V
30 10 V
29 10 V
30 9 V
30 10 V
30 10 V
29 10 V
30 10 V
30 9 V
30 10 V
29 10 V
30 10 V
30 10 V
30 10 V
29 9 V
30 10 V
30 10 V
30 10 V
29 10 V
30 9 V
30 10 V
30 10 V
29 10 V
30 10 V
30 10 V
30 9 V
29 10 V
30 10 V
30 10 V
30 10 V
29 9 V
30 10 V
30 10 V
30 10 V
29 10 V
30 10 V
30 9 V
30 10 V
29 10 V
30 10 V
30 10 V
30 10 V
29 9 V
30 10 V
30 10 V
30 10 V
29 10 V
30 9 V
30 10 V
30 10 V
29 10 V
30 10 V
30 10 V
30 9 V
29 10 V
30 10 V
30 10 V
30 10 V
29 9 V
30 10 V
30 10 V
30 10 V
29 10 V
30 10 V
30 9 V
30 10 V
29 10 V
30 10 V
30 10 V
30 9 V
29 10 V
30 10 V
30 10 V
30 10 V
29 10 V
30 9 V
30 10 V
30 10 V
29 10 V
30 10 V
30 10 V
30 9 V
29 10 V
30 10 V
30 10 V
30 10 V
29 9 V
30 10 V
1.000 UL
LTb
350 300 M
3100 0 V
0 1760 V
-3100 0 V
350 300 L
1.000 UP
stroke
grestore
end
showpage
}}%
\put(1900,50){\makebox(0,0){word length in characters}}%
\put(100,1180){%
\special{ps: gsave currentpoint currentpoint translate
270 rotate neg exch neg exch translate}%
\makebox(0,0)[b]{\shortstack{unpredictability}}%
\special{ps: currentpoint grestore moveto}%
}%
\put(3450,200){\makebox(0,0){ 20}}%
\put(3140,200){\makebox(0,0){ 18}}%
\put(2830,200){\makebox(0,0){ 16}}%
\put(2520,200){\makebox(0,0){ 14}}%
\put(2210,200){\makebox(0,0){ 12}}%
\put(1900,200){\makebox(0,0){ 10}}%
\put(1590,200){\makebox(0,0){ 8}}%
\put(1280,200){\makebox(0,0){ 6}}%
\put(970,200){\makebox(0,0){ 4}}%
\put(660,200){\makebox(0,0){ 2}}%
\put(350,200){\makebox(0,0){ 0}}%
\put(300,2060){\makebox(0,0)[r]{ 7}}%
\put(300,1809){\makebox(0,0)[r]{ 6}}%
\put(300,1557){\makebox(0,0)[r]{ 5}}%
\put(300,1306){\makebox(0,0)[r]{ 4}}%
\put(300,1054){\makebox(0,0)[r]{ 3}}%
\put(300,803){\makebox(0,0)[r]{ 2}}%
\put(300,551){\makebox(0,0)[r]{ 1}}%
\put(300,300){\makebox(0,0)[r]{ 0}}%
\end{picture}%
\endgroup
 

%% file: nepred_all_cyl_e.tex
\begingroup%
  \makeatletter%
  \newcommand{\GNUPLOTspecial}{%
    \@sanitize\catcode`\%=14\relax\special}%
  \setlength{\unitlength}{0.1bp}%
\begin{picture}(3600,2160)(0,0)%
{\GNUPLOTspecial{"
/gnudict 256 dict def
gnudict begin
/Color false def
/Solid false def
/gnulinewidth 5.000 def
/userlinewidth gnulinewidth def
/vshift -33 def
/dl {10.0 mul} def
/hpt_ 31.5 def
/vpt_ 31.5 def
/hpt hpt_ def
/vpt vpt_ def
/Rounded false def
/M {moveto} bind def
/L {lineto} bind def
/R {rmoveto} bind def
/V {rlineto} bind def
/N {newpath moveto} bind def
/C {setrgbcolor} bind def
/f {rlineto fill} bind def
/vpt2 vpt 2 mul def
/hpt2 hpt 2 mul def
/Lshow { currentpoint stroke M
  0 vshift R show } def
/Rshow { currentpoint stroke M
  dup stringwidth pop neg vshift R show } def
/Cshow { currentpoint stroke M
  dup stringwidth pop -2 div vshift R show } def
/UP { dup vpt_ mul /vpt exch def hpt_ mul /hpt exch def
  /hpt2 hpt 2 mul def /vpt2 vpt 2 mul def } def
/DL { Color {setrgbcolor Solid {pop []} if 0 setdash }
 {pop pop pop 0 setgray Solid {pop []} if 0 setdash} ifelse } def
/BL { stroke userlinewidth 2 mul setlinewidth
      Rounded { 1 setlinejoin 1 setlinecap } if } def
/AL { stroke userlinewidth 2 div setlinewidth
      Rounded { 1 setlinejoin 1 setlinecap } if } def
/UL { dup gnulinewidth mul /userlinewidth exch def
      dup 1 lt {pop 1} if 10 mul /udl exch def } def
/PL { stroke userlinewidth setlinewidth
      Rounded { 1 setlinejoin 1 setlinecap } if } def
/LTw { PL [] 1 setgray } def
/LTb { BL [] 0 0 0 DL } def
/LTa { AL [1 udl mul 2 udl mul] 0 setdash 0 0 0 setrgbcolor } def
/LT0 { PL [] 1 0 0 DL } def
/LT1 { PL [4 dl 2 dl] 0 1 0 DL } def
/LT2 { PL [2 dl 3 dl] 0 0 1 DL } def
/LT3 { PL [1 dl 1.5 dl] 1 0 1 DL } def
/LT4 { PL [5 dl 2 dl 1 dl 2 dl] 0 1 1 DL } def
/LT5 { PL [4 dl 3 dl 1 dl 3 dl] 1 1 0 DL } def
/LT6 { PL [2 dl 2 dl 2 dl 4 dl] 0 0 0 DL } def
/LT7 { PL [2 dl 2 dl 2 dl 2 dl 2 dl 4 dl] 1 0.3 0 DL } def
/LT8 { PL [2 dl 2 dl 2 dl 2 dl 2 dl 2 dl 2 dl 4 dl] 0.5 0.5 0.5 DL } def
/Pnt { stroke [] 0 setdash
   gsave 1 setlinecap M 0 0 V stroke grestore } def
/Dia { stroke [] 0 setdash 2 copy vpt add M
  hpt neg vpt neg V hpt vpt neg V
  hpt vpt V hpt neg vpt V closepath stroke
  Pnt } def
/Pls { stroke [] 0 setdash vpt sub M 0 vpt2 V
  currentpoint stroke M
  hpt neg vpt neg R hpt2 0 V stroke
  } def
/Box { stroke [] 0 setdash 2 copy exch hpt sub exch vpt add M
  0 vpt2 neg V hpt2 0 V 0 vpt2 V
  hpt2 neg 0 V closepath stroke
  Pnt } def
/Crs { stroke [] 0 setdash exch hpt sub exch vpt add M
  hpt2 vpt2 neg V currentpoint stroke M
  hpt2 neg 0 R hpt2 vpt2 V stroke } def
/TriU { stroke [] 0 setdash 2 copy vpt 1.12 mul add M
  hpt neg vpt -1.62 mul V
  hpt 2 mul 0 V
  hpt neg vpt 1.62 mul V closepath stroke
  Pnt  } def
/Star { 2 copy Pls Crs } def
/BoxF { stroke [] 0 setdash exch hpt sub exch vpt add M
  0 vpt2 neg V  hpt2 0 V  0 vpt2 V
  hpt2 neg 0 V  closepath fill } def
/TriUF { stroke [] 0 setdash vpt 1.12 mul add M
  hpt neg vpt -1.62 mul V
  hpt 2 mul 0 V
  hpt neg vpt 1.62 mul V closepath fill } def
/TriD { stroke [] 0 setdash 2 copy vpt 1.12 mul sub M
  hpt neg vpt 1.62 mul V
  hpt 2 mul 0 V
  hpt neg vpt -1.62 mul V closepath stroke
  Pnt  } def
/TriDF { stroke [] 0 setdash vpt 1.12 mul sub M
  hpt neg vpt 1.62 mul V
  hpt 2 mul 0 V
  hpt neg vpt -1.62 mul V closepath fill} def
/DiaF { stroke [] 0 setdash vpt add M
  hpt neg vpt neg V hpt vpt neg V
  hpt vpt V hpt neg vpt V closepath fill } def
/Pent { stroke [] 0 setdash 2 copy gsave
  translate 0 hpt M 4 {72 rotate 0 hpt L} repeat
  closepath stroke grestore Pnt } def
/PentF { stroke [] 0 setdash gsave
  translate 0 hpt M 4 {72 rotate 0 hpt L} repeat
  closepath fill grestore } def
/Circle { stroke [] 0 setdash 2 copy
  hpt 0 360 arc stroke Pnt } def
/CircleF { stroke [] 0 setdash hpt 0 360 arc fill } def
/C0 { BL [] 0 setdash 2 copy moveto vpt 90 450  arc } bind def
/C1 { BL [] 0 setdash 2 copy        moveto
       2 copy  vpt 0 90 arc closepath fill
               vpt 0 360 arc closepath } bind def
/C2 { BL [] 0 setdash 2 copy moveto
       2 copy  vpt 90 180 arc closepath fill
               vpt 0 360 arc closepath } bind def
/C3 { BL [] 0 setdash 2 copy moveto
       2 copy  vpt 0 180 arc closepath fill
               vpt 0 360 arc closepath } bind def
/C4 { BL [] 0 setdash 2 copy moveto
       2 copy  vpt 180 270 arc closepath fill
               vpt 0 360 arc closepath } bind def
/C5 { BL [] 0 setdash 2 copy moveto
       2 copy  vpt 0 90 arc
       2 copy moveto
       2 copy  vpt 180 270 arc closepath fill
               vpt 0 360 arc } bind def
/C6 { BL [] 0 setdash 2 copy moveto
      2 copy  vpt 90 270 arc closepath fill
              vpt 0 360 arc closepath } bind def
/C7 { BL [] 0 setdash 2 copy moveto
      2 copy  vpt 0 270 arc closepath fill
              vpt 0 360 arc closepath } bind def
/C8 { BL [] 0 setdash 2 copy moveto
      2 copy vpt 270 360 arc closepath fill
              vpt 0 360 arc closepath } bind def
/C9 { BL [] 0 setdash 2 copy moveto
      2 copy  vpt 270 450 arc closepath fill
              vpt 0 360 arc closepath } bind def
/C10 { BL [] 0 setdash 2 copy 2 copy moveto vpt 270 360 arc closepath fill
       2 copy moveto
       2 copy vpt 90 180 arc closepath fill
               vpt 0 360 arc closepath } bind def
/C11 { BL [] 0 setdash 2 copy moveto
       2 copy  vpt 0 180 arc closepath fill
       2 copy moveto
       2 copy  vpt 270 360 arc closepath fill
               vpt 0 360 arc closepath } bind def
/C12 { BL [] 0 setdash 2 copy moveto
       2 copy  vpt 180 360 arc closepath fill
               vpt 0 360 arc closepath } bind def
/C13 { BL [] 0 setdash  2 copy moveto
       2 copy  vpt 0 90 arc closepath fill
       2 copy moveto
       2 copy  vpt 180 360 arc closepath fill
               vpt 0 360 arc closepath } bind def
/C14 { BL [] 0 setdash 2 copy moveto
       2 copy  vpt 90 360 arc closepath fill
               vpt 0 360 arc } bind def
/C15 { BL [] 0 setdash 2 copy vpt 0 360 arc closepath fill
               vpt 0 360 arc closepath } bind def
/Rec   { newpath 4 2 roll moveto 1 index 0 rlineto 0 exch rlineto
       neg 0 rlineto closepath } bind def
/Square { dup Rec } bind def
/Bsquare { vpt sub exch vpt sub exch vpt2 Square } bind def
/S0 { BL [] 0 setdash 2 copy moveto 0 vpt rlineto BL Bsquare } bind def
/S1 { BL [] 0 setdash 2 copy vpt Square fill Bsquare } bind def
/S2 { BL [] 0 setdash 2 copy exch vpt sub exch vpt Square fill Bsquare } bind def
/S3 { BL [] 0 setdash 2 copy exch vpt sub exch vpt2 vpt Rec fill Bsquare } bind def
/S4 { BL [] 0 setdash 2 copy exch vpt sub exch vpt sub vpt Square fill Bsquare } bind def
/S5 { BL [] 0 setdash 2 copy 2 copy vpt Square fill
       exch vpt sub exch vpt sub vpt Square fill Bsquare } bind def
/S6 { BL [] 0 setdash 2 copy exch vpt sub exch vpt sub vpt vpt2 Rec fill Bsquare } bind def
/S7 { BL [] 0 setdash 2 copy exch vpt sub exch vpt sub vpt vpt2 Rec fill
       2 copy vpt Square fill
       Bsquare } bind def
/S8 { BL [] 0 setdash 2 copy vpt sub vpt Square fill Bsquare } bind def
/S9 { BL [] 0 setdash 2 copy vpt sub vpt vpt2 Rec fill Bsquare } bind def
/S10 { BL [] 0 setdash 2 copy vpt sub vpt Square fill 2 copy exch vpt sub exch vpt Square fill
       Bsquare } bind def
/S11 { BL [] 0 setdash 2 copy vpt sub vpt Square fill 2 copy exch vpt sub exch vpt2 vpt Rec fill
       Bsquare } bind def
/S12 { BL [] 0 setdash 2 copy exch vpt sub exch vpt sub vpt2 vpt Rec fill Bsquare } bind def
/S13 { BL [] 0 setdash 2 copy exch vpt sub exch vpt sub vpt2 vpt Rec fill
       2 copy vpt Square fill Bsquare } bind def
/S14 { BL [] 0 setdash 2 copy exch vpt sub exch vpt sub vpt2 vpt Rec fill
       2 copy exch vpt sub exch vpt Square fill Bsquare } bind def
/S15 { BL [] 0 setdash 2 copy Bsquare fill Bsquare } bind def
/D0 { gsave translate 45 rotate 0 0 S0 stroke grestore } bind def
/D1 { gsave translate 45 rotate 0 0 S1 stroke grestore } bind def
/D2 { gsave translate 45 rotate 0 0 S2 stroke grestore } bind def
/D3 { gsave translate 45 rotate 0 0 S3 stroke grestore } bind def
/D4 { gsave translate 45 rotate 0 0 S4 stroke grestore } bind def
/D5 { gsave translate 45 rotate 0 0 S5 stroke grestore } bind def
/D6 { gsave translate 45 rotate 0 0 S6 stroke grestore } bind def
/D7 { gsave translate 45 rotate 0 0 S7 stroke grestore } bind def
/D8 { gsave translate 45 rotate 0 0 S8 stroke grestore } bind def
/D9 { gsave translate 45 rotate 0 0 S9 stroke grestore } bind def
/D10 { gsave translate 45 rotate 0 0 S10 stroke grestore } bind def
/D11 { gsave translate 45 rotate 0 0 S11 stroke grestore } bind def
/D12 { gsave translate 45 rotate 0 0 S12 stroke grestore } bind def
/D13 { gsave translate 45 rotate 0 0 S13 stroke grestore } bind def
/D14 { gsave translate 45 rotate 0 0 S14 stroke grestore } bind def
/D15 { gsave translate 45 rotate 0 0 S15 stroke grestore } bind def
/DiaE { stroke [] 0 setdash vpt add M
  hpt neg vpt neg V hpt vpt neg V
  hpt vpt V hpt neg vpt V closepath stroke } def
/BoxE { stroke [] 0 setdash exch hpt sub exch vpt add M
  0 vpt2 neg V hpt2 0 V 0 vpt2 V
  hpt2 neg 0 V closepath stroke } def
/TriUE { stroke [] 0 setdash vpt 1.12 mul add M
  hpt neg vpt -1.62 mul V
  hpt 2 mul 0 V
  hpt neg vpt 1.62 mul V closepath stroke } def
/TriDE { stroke [] 0 setdash vpt 1.12 mul sub M
  hpt neg vpt 1.62 mul V
  hpt 2 mul 0 V
  hpt neg vpt -1.62 mul V closepath stroke } def
/PentE { stroke [] 0 setdash gsave
  translate 0 hpt M 4 {72 rotate 0 hpt L} repeat
  closepath stroke grestore } def
/CircE { stroke [] 0 setdash 
  hpt 0 360 arc stroke } def
/Opaque { gsave closepath 1 setgray fill grestore 0 setgray closepath } def
/DiaW { stroke [] 0 setdash vpt add M
  hpt neg vpt neg V hpt vpt neg V
  hpt vpt V hpt neg vpt V Opaque stroke } def
/BoxW { stroke [] 0 setdash exch hpt sub exch vpt add M
  0 vpt2 neg V hpt2 0 V 0 vpt2 V
  hpt2 neg 0 V Opaque stroke } def
/TriUW { stroke [] 0 setdash vpt 1.12 mul add M
  hpt neg vpt -1.62 mul V
  hpt 2 mul 0 V
  hpt neg vpt 1.62 mul V Opaque stroke } def
/TriDW { stroke [] 0 setdash vpt 1.12 mul sub M
  hpt neg vpt 1.62 mul V
  hpt 2 mul 0 V
  hpt neg vpt -1.62 mul V Opaque stroke } def
/PentW { stroke [] 0 setdash gsave
  translate 0 hpt M 4 {72 rotate 0 hpt L} repeat
  Opaque stroke grestore } def
/CircW { stroke [] 0 setdash 
  hpt 0 360 arc Opaque stroke } def
/BoxFill { gsave Rec 1 setgray fill grestore } def
/BoxColFill {
  gsave Rec
  /Fillden exch def
  currentrgbcolor
  /ColB exch def /ColG exch def /ColR exch def
  /ColR ColR Fillden mul Fillden sub 1 add def
  /ColG ColG Fillden mul Fillden sub 1 add def
  /ColB ColB Fillden mul Fillden sub 1 add def
  ColR ColG ColB setrgbcolor
  fill grestore } def
%
%
/PatternFill { gsave /PFa [ 9 2 roll ] def
    PFa 0 get PFa 2 get 2 div add PFa 1 get PFa 3 get 2 div add translate
    PFa 2 get -2 div PFa 3 get -2 div PFa 2 get PFa 3 get Rec
    gsave 1 setgray fill grestore clip
    currentlinewidth 0.5 mul setlinewidth
    /PFs PFa 2 get dup mul PFa 3 get dup mul add sqrt def
    0 0 M PFa 5 get rotate PFs -2 div dup translate
	0 1 PFs PFa 4 get div 1 add floor cvi
	{ PFa 4 get mul 0 M 0 PFs V } for
    0 PFa 6 get ne {
	0 1 PFs PFa 4 get div 1 add floor cvi
	{ PFa 4 get mul 0 2 1 roll M PFs 0 V } for
    } if
    stroke grestore } def
/Symbol-Oblique /Symbol findfont [1 0 .167 1 0 0] makefont
dup length dict begin {1 index /FID eq {pop pop} {def} ifelse} forall
currentdict end definefont pop
end
gnudict begin
gsave
0 0 translate
0.100 0.100 scale
0 setgray
newpath
1.000 UL
LTb
1.000 UL
LTa
450 300 M
3000 0 V
1.000 UL
LTb
450 300 M
63 0 V
2937 0 R
-63 0 V
1.000 UL
LTb
1.000 UL
LTa
450 551 M
3000 0 V
1.000 UL
LTb
450 551 M
63 0 V
2937 0 R
-63 0 V
1.000 UL
LTb
1.000 UL
LTa
450 803 M
3000 0 V
1.000 UL
LTb
450 803 M
63 0 V
2937 0 R
-63 0 V
1.000 UL
LTb
1.000 UL
LTa
450 1054 M
3000 0 V
1.000 UL
LTb
450 1054 M
63 0 V
2937 0 R
-63 0 V
1.000 UL
LTb
1.000 UL
LTa
450 1306 M
3000 0 V
1.000 UL
LTb
450 1306 M
63 0 V
2937 0 R
-63 0 V
1.000 UL
LTb
1.000 UL
LTa
450 1557 M
3000 0 V
1.000 UL
LTb
450 1557 M
63 0 V
2937 0 R
-63 0 V
1.000 UL
LTb
1.000 UL
LTa
450 1809 M
3000 0 V
1.000 UL
LTb
450 1809 M
63 0 V
2937 0 R
-63 0 V
1.000 UL
LTb
1.000 UL
LTa
450 2060 M
3000 0 V
1.000 UL
LTb
450 2060 M
63 0 V
2937 0 R
-63 0 V
1.000 UL
LTb
1.000 UL
LTa
450 300 M
0 1760 V
1.000 UL
LTb
450 300 M
0 63 V
0 1697 R
0 -63 V
1.000 UL
LTb
1.000 UL
LTa
879 300 M
0 1760 V
1.000 UL
LTb
879 300 M
0 63 V
0 1697 R
0 -63 V
1.000 UL
LTb
1.000 UL
LTa
1307 300 M
0 1760 V
1.000 UL
LTb
1307 300 M
0 63 V
0 1697 R
0 -63 V
1.000 UL
LTb
1.000 UL
LTa
1736 300 M
0 1760 V
1.000 UL
LTb
1736 300 M
0 63 V
0 1697 R
0 -63 V
1.000 UL
LTb
1.000 UL
LTa
2164 300 M
0 1760 V
1.000 UL
LTb
2164 300 M
0 63 V
0 1697 R
0 -63 V
1.000 UL
LTb
1.000 UL
LTa
2593 300 M
0 1760 V
1.000 UL
LTb
2593 300 M
0 63 V
0 1697 R
0 -63 V
1.000 UL
LTb
1.000 UL
LTa
3021 300 M
0 1760 V
1.000 UL
LTb
3021 300 M
0 63 V
0 1697 R
0 -63 V
1.000 UL
LTb
1.000 UL
LTa
3450 300 M
0 1760 V
1.000 UL
LTb
3450 300 M
0 63 V
0 1697 R
0 -63 V
1.000 UL
LTb
1.000 UL
LTb
450 300 M
3000 0 V
0 1760 V
-3000 0 V
450 300 L
LTb
LTb
1.000 UP
1.000 UP
1.000 UL
LT0
879 733 M
0 20 V
848 733 M
62 0 V
-62 20 R
62 0 V
397 131 R
0 10 V
-31 -10 R
62 0 V
-62 10 R
62 0 V
398 197 R
0 10 V
-31 -10 R
62 0 V
-62 10 R
62 0 V
397 197 R
0 20 V
-31 -20 R
62 0 V
-62 20 R
62 0 V
398 193 R
0 40 V
-31 -40 R
62 0 V
-62 40 R
62 0 V
397 -31 R
0 90 V
-31 -90 R
62 0 V
-62 90 R
62 0 V
398 -149 R
0 191 V
-31 -191 R
62 0 V
-62 191 R
62 0 V
879 743 Pls
1307 889 Pls
1736 1096 Pls
2164 1308 Pls
2593 1531 Pls
3021 1565 Pls
3450 1556 Pls
1.000 UL
LT1
450 515 M
30 14 V
31 14 V
30 14 V
30 14 V
31 14 V
30 15 V
30 14 V
30 14 V
31 14 V
30 14 V
30 14 V
31 14 V
30 14 V
30 14 V
31 15 V
30 14 V
30 14 V
30 14 V
31 14 V
30 14 V
30 14 V
31 14 V
30 14 V
30 15 V
31 14 V
30 14 V
30 14 V
30 14 V
31 14 V
30 14 V
30 14 V
31 14 V
30 14 V
30 15 V
31 14 V
30 14 V
30 14 V
31 14 V
30 14 V
30 14 V
30 14 V
31 14 V
30 15 V
30 14 V
31 14 V
30 14 V
30 14 V
31 14 V
30 14 V
30 14 V
30 14 V
31 15 V
30 14 V
30 14 V
31 14 V
30 14 V
30 14 V
31 14 V
30 14 V
30 14 V
30 15 V
31 14 V
30 14 V
30 14 V
31 14 V
30 14 V
30 14 V
31 14 V
30 14 V
30 15 V
31 14 V
30 14 V
30 14 V
30 14 V
31 14 V
30 14 V
30 14 V
31 14 V
30 14 V
30 15 V
31 14 V
30 14 V
30 14 V
30 14 V
31 14 V
30 14 V
30 14 V
31 14 V
30 15 V
30 14 V
31 14 V
30 14 V
30 14 V
30 14 V
31 14 V
30 14 V
30 14 V
31 15 V
30 14 V
1.000 UL
LTb
450 300 M
3000 0 V
0 1760 V
-3000 0 V
450 300 L
1.000 UP
stroke
grestore
end
showpage
}}%
\put(1950,50){\makebox(0,0){word length in cyllables}}%
\put(100,1180){%
\special{ps: gsave currentpoint currentpoint translate
270 rotate neg exch neg exch translate}%
\makebox(0,0)[b]{\shortstack{unpredictability}}%
\special{ps: currentpoint grestore moveto}%
}%
\put(3450,200){\makebox(0,0){ 7}}%
\put(3021,200){\makebox(0,0){ 6}}%
\put(2593,200){\makebox(0,0){ 5}}%
\put(2164,200){\makebox(0,0){ 4}}%
\put(1736,200){\makebox(0,0){ 3}}%
\put(1307,200){\makebox(0,0){ 2}}%
\put(879,200){\makebox(0,0){ 1}}%
\put(450,200){\makebox(0,0){ 0}}%
\put(400,2060){\makebox(0,0)[r]{ 3.5}}%
\put(400,1809){\makebox(0,0)[r]{ 3}}%
\put(400,1557){\makebox(0,0)[r]{ 2.5}}%
\put(400,1306){\makebox(0,0)[r]{ 2}}%
\put(400,1054){\makebox(0,0)[r]{ 1.5}}%
\put(400,803){\makebox(0,0)[r]{ 1}}%
\put(400,551){\makebox(0,0)[r]{ 0.5}}%
\put(400,300){\makebox(0,0)[r]{ 0}}%
\end{picture}%
\endgroup
 

%% file: nepred_prose_let_e.tex
\begingroup%
  \makeatletter%
  \newcommand{\GNUPLOTspecial}{%
    \@sanitize\catcode`\%=14\relax\special}%
  \setlength{\unitlength}{0.1bp}%
\begin{picture}(3600,2160)(0,0)%
{\GNUPLOTspecial{"
/gnudict 256 dict def
gnudict begin
/Color false def
/Solid false def
/gnulinewidth 5.000 def
/userlinewidth gnulinewidth def
/vshift -33 def
/dl {10.0 mul} def
/hpt_ 31.5 def
/vpt_ 31.5 def
/hpt hpt_ def
/vpt vpt_ def
/Rounded false def
/M {moveto} bind def
/L {lineto} bind def
/R {rmoveto} bind def
/V {rlineto} bind def
/N {newpath moveto} bind def
/C {setrgbcolor} bind def
/f {rlineto fill} bind def
/vpt2 vpt 2 mul def
/hpt2 hpt 2 mul def
/Lshow { currentpoint stroke M
  0 vshift R show } def
/Rshow { currentpoint stroke M
  dup stringwidth pop neg vshift R show } def
/Cshow { currentpoint stroke M
  dup stringwidth pop -2 div vshift R show } def
/UP { dup vpt_ mul /vpt exch def hpt_ mul /hpt exch def
  /hpt2 hpt 2 mul def /vpt2 vpt 2 mul def } def
/DL { Color {setrgbcolor Solid {pop []} if 0 setdash }
 {pop pop pop 0 setgray Solid {pop []} if 0 setdash} ifelse } def
/BL { stroke userlinewidth 2 mul setlinewidth
      Rounded { 1 setlinejoin 1 setlinecap } if } def
/AL { stroke userlinewidth 2 div setlinewidth
      Rounded { 1 setlinejoin 1 setlinecap } if } def
/UL { dup gnulinewidth mul /userlinewidth exch def
      dup 1 lt {pop 1} if 10 mul /udl exch def } def
/PL { stroke userlinewidth setlinewidth
      Rounded { 1 setlinejoin 1 setlinecap } if } def
/LTw { PL [] 1 setgray } def
/LTb { BL [] 0 0 0 DL } def
/LTa { AL [1 udl mul 2 udl mul] 0 setdash 0 0 0 setrgbcolor } def
/LT0 { PL [] 1 0 0 DL } def
/LT1 { PL [4 dl 2 dl] 0 1 0 DL } def
/LT2 { PL [2 dl 3 dl] 0 0 1 DL } def
/LT3 { PL [1 dl 1.5 dl] 1 0 1 DL } def
/LT4 { PL [5 dl 2 dl 1 dl 2 dl] 0 1 1 DL } def
/LT5 { PL [4 dl 3 dl 1 dl 3 dl] 1 1 0 DL } def
/LT6 { PL [2 dl 2 dl 2 dl 4 dl] 0 0 0 DL } def
/LT7 { PL [2 dl 2 dl 2 dl 2 dl 2 dl 4 dl] 1 0.3 0 DL } def
/LT8 { PL [2 dl 2 dl 2 dl 2 dl 2 dl 2 dl 2 dl 4 dl] 0.5 0.5 0.5 DL } def
/Pnt { stroke [] 0 setdash
   gsave 1 setlinecap M 0 0 V stroke grestore } def
/Dia { stroke [] 0 setdash 2 copy vpt add M
  hpt neg vpt neg V hpt vpt neg V
  hpt vpt V hpt neg vpt V closepath stroke
  Pnt } def
/Pls { stroke [] 0 setdash vpt sub M 0 vpt2 V
  currentpoint stroke M
  hpt neg vpt neg R hpt2 0 V stroke
  } def
/Box { stroke [] 0 setdash 2 copy exch hpt sub exch vpt add M
  0 vpt2 neg V hpt2 0 V 0 vpt2 V
  hpt2 neg 0 V closepath stroke
  Pnt } def
/Crs { stroke [] 0 setdash exch hpt sub exch vpt add M
  hpt2 vpt2 neg V currentpoint stroke M
  hpt2 neg 0 R hpt2 vpt2 V stroke } def
/TriU { stroke [] 0 setdash 2 copy vpt 1.12 mul add M
  hpt neg vpt -1.62 mul V
  hpt 2 mul 0 V
  hpt neg vpt 1.62 mul V closepath stroke
  Pnt  } def
/Star { 2 copy Pls Crs } def
/BoxF { stroke [] 0 setdash exch hpt sub exch vpt add M
  0 vpt2 neg V  hpt2 0 V  0 vpt2 V
  hpt2 neg 0 V  closepath fill } def
/TriUF { stroke [] 0 setdash vpt 1.12 mul add M
  hpt neg vpt -1.62 mul V
  hpt 2 mul 0 V
  hpt neg vpt 1.62 mul V closepath fill } def
/TriD { stroke [] 0 setdash 2 copy vpt 1.12 mul sub M
  hpt neg vpt 1.62 mul V
  hpt 2 mul 0 V
  hpt neg vpt -1.62 mul V closepath stroke
  Pnt  } def
/TriDF { stroke [] 0 setdash vpt 1.12 mul sub M
  hpt neg vpt 1.62 mul V
  hpt 2 mul 0 V
  hpt neg vpt -1.62 mul V closepath fill} def
/DiaF { stroke [] 0 setdash vpt add M
  hpt neg vpt neg V hpt vpt neg V
  hpt vpt V hpt neg vpt V closepath fill } def
/Pent { stroke [] 0 setdash 2 copy gsave
  translate 0 hpt M 4 {72 rotate 0 hpt L} repeat
  closepath stroke grestore Pnt } def
/PentF { stroke [] 0 setdash gsave
  translate 0 hpt M 4 {72 rotate 0 hpt L} repeat
  closepath fill grestore } def
/Circle { stroke [] 0 setdash 2 copy
  hpt 0 360 arc stroke Pnt } def
/CircleF { stroke [] 0 setdash hpt 0 360 arc fill } def
/C0 { BL [] 0 setdash 2 copy moveto vpt 90 450  arc } bind def
/C1 { BL [] 0 setdash 2 copy        moveto
       2 copy  vpt 0 90 arc closepath fill
               vpt 0 360 arc closepath } bind def
/C2 { BL [] 0 setdash 2 copy moveto
       2 copy  vpt 90 180 arc closepath fill
               vpt 0 360 arc closepath } bind def
/C3 { BL [] 0 setdash 2 copy moveto
       2 copy  vpt 0 180 arc closepath fill
               vpt 0 360 arc closepath } bind def
/C4 { BL [] 0 setdash 2 copy moveto
       2 copy  vpt 180 270 arc closepath fill
               vpt 0 360 arc closepath } bind def
/C5 { BL [] 0 setdash 2 copy moveto
       2 copy  vpt 0 90 arc
       2 copy moveto
       2 copy  vpt 180 270 arc closepath fill
               vpt 0 360 arc } bind def
/C6 { BL [] 0 setdash 2 copy moveto
      2 copy  vpt 90 270 arc closepath fill
              vpt 0 360 arc closepath } bind def
/C7 { BL [] 0 setdash 2 copy moveto
      2 copy  vpt 0 270 arc closepath fill
              vpt 0 360 arc closepath } bind def
/C8 { BL [] 0 setdash 2 copy moveto
      2 copy vpt 270 360 arc closepath fill
              vpt 0 360 arc closepath } bind def
/C9 { BL [] 0 setdash 2 copy moveto
      2 copy  vpt 270 450 arc closepath fill
              vpt 0 360 arc closepath } bind def
/C10 { BL [] 0 setdash 2 copy 2 copy moveto vpt 270 360 arc closepath fill
       2 copy moveto
       2 copy vpt 90 180 arc closepath fill
               vpt 0 360 arc closepath } bind def
/C11 { BL [] 0 setdash 2 copy moveto
       2 copy  vpt 0 180 arc closepath fill
       2 copy moveto
       2 copy  vpt 270 360 arc closepath fill
               vpt 0 360 arc closepath } bind def
/C12 { BL [] 0 setdash 2 copy moveto
       2 copy  vpt 180 360 arc closepath fill
               vpt 0 360 arc closepath } bind def
/C13 { BL [] 0 setdash  2 copy moveto
       2 copy  vpt 0 90 arc closepath fill
       2 copy moveto
       2 copy  vpt 180 360 arc closepath fill
               vpt 0 360 arc closepath } bind def
/C14 { BL [] 0 setdash 2 copy moveto
       2 copy  vpt 90 360 arc closepath fill
               vpt 0 360 arc } bind def
/C15 { BL [] 0 setdash 2 copy vpt 0 360 arc closepath fill
               vpt 0 360 arc closepath } bind def
/Rec   { newpath 4 2 roll moveto 1 index 0 rlineto 0 exch rlineto
       neg 0 rlineto closepath } bind def
/Square { dup Rec } bind def
/Bsquare { vpt sub exch vpt sub exch vpt2 Square } bind def
/S0 { BL [] 0 setdash 2 copy moveto 0 vpt rlineto BL Bsquare } bind def
/S1 { BL [] 0 setdash 2 copy vpt Square fill Bsquare } bind def
/S2 { BL [] 0 setdash 2 copy exch vpt sub exch vpt Square fill Bsquare } bind def
/S3 { BL [] 0 setdash 2 copy exch vpt sub exch vpt2 vpt Rec fill Bsquare } bind def
/S4 { BL [] 0 setdash 2 copy exch vpt sub exch vpt sub vpt Square fill Bsquare } bind def
/S5 { BL [] 0 setdash 2 copy 2 copy vpt Square fill
       exch vpt sub exch vpt sub vpt Square fill Bsquare } bind def
/S6 { BL [] 0 setdash 2 copy exch vpt sub exch vpt sub vpt vpt2 Rec fill Bsquare } bind def
/S7 { BL [] 0 setdash 2 copy exch vpt sub exch vpt sub vpt vpt2 Rec fill
       2 copy vpt Square fill
       Bsquare } bind def
/S8 { BL [] 0 setdash 2 copy vpt sub vpt Square fill Bsquare } bind def
/S9 { BL [] 0 setdash 2 copy vpt sub vpt vpt2 Rec fill Bsquare } bind def
/S10 { BL [] 0 setdash 2 copy vpt sub vpt Square fill 2 copy exch vpt sub exch vpt Square fill
       Bsquare } bind def
/S11 { BL [] 0 setdash 2 copy vpt sub vpt Square fill 2 copy exch vpt sub exch vpt2 vpt Rec fill
       Bsquare } bind def
/S12 { BL [] 0 setdash 2 copy exch vpt sub exch vpt sub vpt2 vpt Rec fill Bsquare } bind def
/S13 { BL [] 0 setdash 2 copy exch vpt sub exch vpt sub vpt2 vpt Rec fill
       2 copy vpt Square fill Bsquare } bind def
/S14 { BL [] 0 setdash 2 copy exch vpt sub exch vpt sub vpt2 vpt Rec fill
       2 copy exch vpt sub exch vpt Square fill Bsquare } bind def
/S15 { BL [] 0 setdash 2 copy Bsquare fill Bsquare } bind def
/D0 { gsave translate 45 rotate 0 0 S0 stroke grestore } bind def
/D1 { gsave translate 45 rotate 0 0 S1 stroke grestore } bind def
/D2 { gsave translate 45 rotate 0 0 S2 stroke grestore } bind def
/D3 { gsave translate 45 rotate 0 0 S3 stroke grestore } bind def
/D4 { gsave translate 45 rotate 0 0 S4 stroke grestore } bind def
/D5 { gsave translate 45 rotate 0 0 S5 stroke grestore } bind def
/D6 { gsave translate 45 rotate 0 0 S6 stroke grestore } bind def
/D7 { gsave translate 45 rotate 0 0 S7 stroke grestore } bind def
/D8 { gsave translate 45 rotate 0 0 S8 stroke grestore } bind def
/D9 { gsave translate 45 rotate 0 0 S9 stroke grestore } bind def
/D10 { gsave translate 45 rotate 0 0 S10 stroke grestore } bind def
/D11 { gsave translate 45 rotate 0 0 S11 stroke grestore } bind def
/D12 { gsave translate 45 rotate 0 0 S12 stroke grestore } bind def
/D13 { gsave translate 45 rotate 0 0 S13 stroke grestore } bind def
/D14 { gsave translate 45 rotate 0 0 S14 stroke grestore } bind def
/D15 { gsave translate 45 rotate 0 0 S15 stroke grestore } bind def
/DiaE { stroke [] 0 setdash vpt add M
  hpt neg vpt neg V hpt vpt neg V
  hpt vpt V hpt neg vpt V closepath stroke } def
/BoxE { stroke [] 0 setdash exch hpt sub exch vpt add M
  0 vpt2 neg V hpt2 0 V 0 vpt2 V
  hpt2 neg 0 V closepath stroke } def
/TriUE { stroke [] 0 setdash vpt 1.12 mul add M
  hpt neg vpt -1.62 mul V
  hpt 2 mul 0 V
  hpt neg vpt 1.62 mul V closepath stroke } def
/TriDE { stroke [] 0 setdash vpt 1.12 mul sub M
  hpt neg vpt 1.62 mul V
  hpt 2 mul 0 V
  hpt neg vpt -1.62 mul V closepath stroke } def
/PentE { stroke [] 0 setdash gsave
  translate 0 hpt M 4 {72 rotate 0 hpt L} repeat
  closepath stroke grestore } def
/CircE { stroke [] 0 setdash 
  hpt 0 360 arc stroke } def
/Opaque { gsave closepath 1 setgray fill grestore 0 setgray closepath } def
/DiaW { stroke [] 0 setdash vpt add M
  hpt neg vpt neg V hpt vpt neg V
  hpt vpt V hpt neg vpt V Opaque stroke } def
/BoxW { stroke [] 0 setdash exch hpt sub exch vpt add M
  0 vpt2 neg V hpt2 0 V 0 vpt2 V
  hpt2 neg 0 V Opaque stroke } def
/TriUW { stroke [] 0 setdash vpt 1.12 mul add M
  hpt neg vpt -1.62 mul V
  hpt 2 mul 0 V
  hpt neg vpt 1.62 mul V Opaque stroke } def
/TriDW { stroke [] 0 setdash vpt 1.12 mul sub M
  hpt neg vpt 1.62 mul V
  hpt 2 mul 0 V
  hpt neg vpt -1.62 mul V Opaque stroke } def
/PentW { stroke [] 0 setdash gsave
  translate 0 hpt M 4 {72 rotate 0 hpt L} repeat
  Opaque stroke grestore } def
/CircW { stroke [] 0 setdash 
  hpt 0 360 arc Opaque stroke } def
/BoxFill { gsave Rec 1 setgray fill grestore } def
/BoxColFill {
  gsave Rec
  /Fillden exch def
  currentrgbcolor
  /ColB exch def /ColG exch def /ColR exch def
  /ColR ColR Fillden mul Fillden sub 1 add def
  /ColG ColG Fillden mul Fillden sub 1 add def
  /ColB ColB Fillden mul Fillden sub 1 add def
  ColR ColG ColB setrgbcolor
  fill grestore } def
%
%
/PatternFill { gsave /PFa [ 9 2 roll ] def
    PFa 0 get PFa 2 get 2 div add PFa 1 get PFa 3 get 2 div add translate
    PFa 2 get -2 div PFa 3 get -2 div PFa 2 get PFa 3 get Rec
    gsave 1 setgray fill grestore clip
    currentlinewidth 0.5 mul setlinewidth
    /PFs PFa 2 get dup mul PFa 3 get dup mul add sqrt def
    0 0 M PFa 5 get rotate PFs -2 div dup translate
	0 1 PFs PFa 4 get div 1 add floor cvi
	{ PFa 4 get mul 0 M 0 PFs V } for
    0 PFa 6 get ne {
	0 1 PFs PFa 4 get div 1 add floor cvi
	{ PFa 4 get mul 0 2 1 roll M PFs 0 V } for
    } if
    stroke grestore } def
/Symbol-Oblique /Symbol findfont [1 0 .167 1 0 0] makefont
dup length dict begin {1 index /FID eq {pop pop} {def} ifelse} forall
currentdict end definefont pop
end
gnudict begin
gsave
0 0 translate
0.100 0.100 scale
0 setgray
newpath
1.000 UL
LTb
1.000 UL
LTa
350 300 M
3100 0 V
1.000 UL
LTb
350 300 M
63 0 V
3037 0 R
-63 0 V
1.000 UL
LTb
1.000 UL
LTa
350 551 M
3100 0 V
1.000 UL
LTb
350 551 M
63 0 V
3037 0 R
-63 0 V
1.000 UL
LTb
1.000 UL
LTa
350 803 M
3100 0 V
1.000 UL
LTb
350 803 M
63 0 V
3037 0 R
-63 0 V
1.000 UL
LTb
1.000 UL
LTa
350 1054 M
3100 0 V
1.000 UL
LTb
350 1054 M
63 0 V
3037 0 R
-63 0 V
1.000 UL
LTb
1.000 UL
LTa
350 1306 M
3100 0 V
1.000 UL
LTb
350 1306 M
63 0 V
3037 0 R
-63 0 V
1.000 UL
LTb
1.000 UL
LTa
350 1557 M
3100 0 V
1.000 UL
LTb
350 1557 M
63 0 V
3037 0 R
-63 0 V
1.000 UL
LTb
1.000 UL
LTa
350 1809 M
3100 0 V
1.000 UL
LTb
350 1809 M
63 0 V
3037 0 R
-63 0 V
1.000 UL
LTb
1.000 UL
LTa
350 2060 M
3100 0 V
1.000 UL
LTb
350 2060 M
63 0 V
3037 0 R
-63 0 V
1.000 UL
LTb
1.000 UL
LTa
350 300 M
0 1760 V
1.000 UL
LTb
350 300 M
0 63 V
0 1697 R
0 -63 V
1.000 UL
LTb
1.000 UL
LTa
660 300 M
0 1760 V
1.000 UL
LTb
660 300 M
0 63 V
0 1697 R
0 -63 V
1.000 UL
LTb
1.000 UL
LTa
970 300 M
0 1760 V
1.000 UL
LTb
970 300 M
0 63 V
0 1697 R
0 -63 V
1.000 UL
LTb
1.000 UL
LTa
1280 300 M
0 1760 V
1.000 UL
LTb
1280 300 M
0 63 V
0 1697 R
0 -63 V
1.000 UL
LTb
1.000 UL
LTa
1590 300 M
0 1760 V
1.000 UL
LTb
1590 300 M
0 63 V
0 1697 R
0 -63 V
1.000 UL
LTb
1.000 UL
LTa
1900 300 M
0 1760 V
1.000 UL
LTb
1900 300 M
0 63 V
0 1697 R
0 -63 V
1.000 UL
LTb
1.000 UL
LTa
2210 300 M
0 1760 V
1.000 UL
LTb
2210 300 M
0 63 V
0 1697 R
0 -63 V
1.000 UL
LTb
1.000 UL
LTa
2520 300 M
0 1760 V
1.000 UL
LTb
2520 300 M
0 63 V
0 1697 R
0 -63 V
1.000 UL
LTb
1.000 UL
LTa
2830 300 M
0 1760 V
1.000 UL
LTb
2830 300 M
0 63 V
0 1697 R
0 -63 V
1.000 UL
LTb
1.000 UL
LTa
3140 300 M
0 1760 V
1.000 UL
LTb
3140 300 M
0 63 V
0 1697 R
0 -63 V
1.000 UL
LTb
1.000 UL
LTa
3450 300 M
0 1760 V
1.000 UL
LTb
3450 300 M
0 63 V
0 1697 R
0 -63 V
1.000 UL
LTb
1.000 UL
LTb
350 300 M
3100 0 V
0 1760 V
-3100 0 V
350 300 L
LTb
LTb
1.000 UP
1.000 UP
1.000 UL
LT0
1125 565 M
0 10 V
-31 -10 R
62 0 V
-62 10 R
62 0 V
124 67 R
0 15 V
-31 -15 R
62 0 V
-62 15 R
62 0 V
124 -1 R
0 15 V
-31 -15 R
62 0 V
-62 15 R
62 0 V
124 58 R
0 20 V
-31 -20 R
62 0 V
-62 20 R
62 0 V
124 19 R
0 25 V
-31 -25 R
62 0 V
-62 25 R
62 0 V
124 31 R
0 36 V
-31 -36 R
62 0 V
-62 36 R
62 0 V
124 -65 R
0 35 V
-31 -35 R
62 0 V
-62 35 R
62 0 V
124 35 R
0 50 V
-31 -50 R
62 0 V
-62 50 R
62 0 V
124 29 R
0 70 V
-31 -70 R
62 0 V
-62 70 R
62 0 V
124 -97 R
0 96 V
-31 -96 R
62 0 V
-62 96 R
62 0 V
2675 911 M
0 121 V
2644 911 M
62 0 V
-62 121 R
62 0 V
2830 711 M
0 101 V
2799 711 M
62 0 V
-62 101 R
62 0 V
124 370 R
0 352 V
-31 -352 R
62 0 V
-62 352 R
62 0 V
124 -370 R
0 714 V
-31 -714 R
62 0 V
-62 714 R
62 0 V
3295 942 M
0 488 V
3264 942 M
62 0 V
-62 488 R
62 0 V
1125 570 Pls
1280 649 Pls
1435 663 Pls
1590 739 Pls
1745 780 Pls
1900 842 Pls
2055 813 Pls
2210 890 Pls
2365 979 Pls
2520 965 Pls
2675 971 Pls
2830 761 Pls
2985 1358 Pls
3140 1521 Pls
3295 1186 Pls
1.000 UL
LT1
350 370 M
30 8 V
29 9 V
30 8 V
30 9 V
30 8 V
29 9 V
30 8 V
30 9 V
30 8 V
29 9 V
30 8 V
30 9 V
30 8 V
29 9 V
30 8 V
30 8 V
30 9 V
29 8 V
30 9 V
30 8 V
30 9 V
29 8 V
30 9 V
30 8 V
30 9 V
29 8 V
30 9 V
30 8 V
30 9 V
29 8 V
30 8 V
30 9 V
30 8 V
29 9 V
30 8 V
30 9 V
30 8 V
29 9 V
30 8 V
30 9 V
30 8 V
29 9 V
30 8 V
30 9 V
30 8 V
29 8 V
30 9 V
30 8 V
30 9 V
29 8 V
30 9 V
30 8 V
30 9 V
29 8 V
30 9 V
30 8 V
30 9 V
29 8 V
30 9 V
30 8 V
30 8 V
29 9 V
30 8 V
30 9 V
30 8 V
29 9 V
30 8 V
30 9 V
30 8 V
29 9 V
30 8 V
30 9 V
30 8 V
29 9 V
30 8 V
30 8 V
30 9 V
29 8 V
30 9 V
30 8 V
30 9 V
29 8 V
30 9 V
30 8 V
30 9 V
29 8 V
30 9 V
30 8 V
30 9 V
29 8 V
30 8 V
30 9 V
30 8 V
29 9 V
30 8 V
30 9 V
30 8 V
29 9 V
30 8 V
1.000 UL
LTb
350 300 M
3100 0 V
0 1760 V
-3100 0 V
350 300 L
1.000 UP
stroke
grestore
end
showpage
}}%
\put(1900,50){\makebox(0,0){word length in characters}}%
\put(100,1180){%
\special{ps: gsave currentpoint currentpoint translate
270 rotate neg exch neg exch translate}%
\makebox(0,0)[b]{\shortstack{unpredictability}}%
\special{ps: currentpoint grestore moveto}%
}%
\put(3450,200){\makebox(0,0){ 20}}%
\put(3140,200){\makebox(0,0){ 18}}%
\put(2830,200){\makebox(0,0){ 16}}%
\put(2520,200){\makebox(0,0){ 14}}%
\put(2210,200){\makebox(0,0){ 12}}%
\put(1900,200){\makebox(0,0){ 10}}%
\put(1590,200){\makebox(0,0){ 8}}%
\put(1280,200){\makebox(0,0){ 6}}%
\put(970,200){\makebox(0,0){ 4}}%
\put(660,200){\makebox(0,0){ 2}}%
\put(350,200){\makebox(0,0){ 0}}%
\put(300,2060){\makebox(0,0)[r]{ 7}}%
\put(300,1809){\makebox(0,0)[r]{ 6}}%
\put(300,1557){\makebox(0,0)[r]{ 5}}%
\put(300,1306){\makebox(0,0)[r]{ 4}}%
\put(300,1054){\makebox(0,0)[r]{ 3}}%
\put(300,803){\makebox(0,0)[r]{ 2}}%
\put(300,551){\makebox(0,0)[r]{ 1}}%
\put(300,300){\makebox(0,0)[r]{ 0}}%
\end{picture}%
\endgroup
 